\documentstyle[11pt,epsf,psfig]{article}

\begin{document}
\begin{titlepage}   

\begin{center}

\vskip0.9truecm
{\bf THE GAMMA-RAY BURST RATE AT HIGH PHOTON ENERGIES}
{\footnote{
To appear in The Astrophysical Journal,  Vol. 467, Aug 20 issue (1996)}}

\vskip 0.5truecm

Karl Mannheim,$^1$ Dieter Hartmann,$^2$ and Burkhardt Funk$^3$\\
\smallskip
{$^1$Universit\"ats-Sternwarte, Geismarlandstrasse 11,
D-37803 G\"ottingen, Germany; kmannhe@uni-sw.gwdg.de.}\\
{$^2$Deptartment of Physics and Astronomy, Clemson University, 
Clemson, SC 29634, USA; hartmann@grb.phys.clemson.edu.}\\
{$^3$Universit\"at Wuppertal, Fachbereich Physik, D-42097 Wuppertal, 
Germany; funk@wpos1.physik.uni-wuppertal.de.}\\

\end{center}

\begin{abstract}
Some gamma-ray burst (GRB) spectra exhibit high 
energy tails with the highest photon energy detected at
$18$~GeV. The spectral slope of the high-energy tails is
sufficiently flat in $\nu F_\nu$ to consider the possibility
of their detection at still higher energies.  
We calculate how many bursts can
reasonably be expected above a given energy threshold for
a cosmological distribution of bursts satisfying the observed
apparent brightness distribution.
The crucial
point is that the gamma-ray absorption by pair production in
the intergalactic diffuse radiation
field eliminates bursts from beyond the
{\it gamma-ray horizon} $\tau_{\gamma\gamma}\sim 1$,
thus drastically reducing the number of bursts at high
energies.
Our results are consistent with the non-detection of bursts by 
current experiments in the 100\ GeV to 100\ TeV energy range. 
For the earth-bound detector array
MILAGRO, we predict a maximal GRB rate of $\sim 10$ events per year.
The Whipple Observatory can detect, under favorable conditions, 
$\sim 1 $  event per year.
The event rate for the HEGRA array  
is $\sim 0.01$ per year.
Detection of significantly higher rates of bursts would
severely challenge cosmological burst scenarios. 
\end{abstract} 

\vskip1cm
\noindent
{\it Subject Headings:} 

\noindent Cosmology: Diffuse Radiation,
Distance Scale, Intergalactic Medium;
Gamma Rays: Bursts
\end{titlepage}
\newpage


\section{Introduction}
The lack of a significant large scale anisotropy in the 
angular distribution of gamma-ray bursts (GRBs) argues
in favor of their cosmological origin (e.g., Meegan {\it et al.} 1995; Briggs
{\it et al.} 1996; Tegmark {\it et al.} 1996). Assuming they are
all standard candles, the observed number of bursts at a
given flux relates directly to a number of sources
at a given redshift. The maximum redshift sampled by BATSE
under these assumptions is $z_{\rm max}\sim 2$ (e.g., Cohen \& Piran 1995).  
Evolution of the burst population can modify this value.
If bursts would happen to be more active in 
recent cosmic history, the maximum redshift would be lower.
However, the maximum
redshift cannot be much lower than $\sim$ 1 in any scenario in which
the bursts trace the distribution of galaxies.
Galaxy clusters are known to be concentrated toward
the supergalactic plane (e.g., Tully 1992; Kolatt, Dekel, $\&$
Lahav 1995).   The brightest (and thus nearest) bursts would reflect
the mass concentrations of the nearby local universe.  However,
no significant deviation from isotropy
has been found in the 3B-catalog data corrected for sky
exposure (Hartmann, Briggs, $\&$ Mannheim 1996).

If GRBs are indeed
seen to a redshift of $\sim$2, absorption of the high 
energy gamma-rays by diffuse background radiation must
become important at photon energies greater than 30\, GeV.  
Since
gamma-ray emitting blazars are also subject to pair absorption,
and since their redshifts are often accurately known, one can 
determine the density of the diffuse radiation
fields from the infrared to the UV by measuring their gamma-ray
cutoff energies (Stecker, de Jager, \& Salamon 1992;
de Jager, Stecker, \& Salamon  1994; Dwek \& Slavin 1994; Mannheim
{\it et al.} 1996; Madau \& Phinney 1996). 
For blazars, detailed spectral models exist which can discriminate
between internal and external absorption 
(e.g., Blandford \& Levinson 1995; Mannheim 1993).

In fact, one could calibrate the distance scale
to gamma-ray bursts by measuring their cutoff energies
due to external absorption using the background radiation density
inferred from blazar observations. For such a program to be
successful, nature must provide 
burst spectra reaching very high photon
energies. Currently, our knowledge about high-energy
burst spectra is rather sparse, and it is not known whether
they exhibit an intrinsic
turnover in the 10\,GeV\,$\sim$\,30\,TeV region or not (Hurley 1996).
However, burst statistics are consistent with all BATSE bursts having
high-energy tails such as the ones observed by EGRET (Dingus {\it et al.} 1995).  
Some of the
bright GRBs which received exposure by EGRET were detected at
GeV energies without evidence for a spectral rollover (Hurley {\it et al.}
1994).
This raises the hope that the {\it intrinsic} spectra continue to still higher
energies and that one could in turn use the expected exponential turnover
to set a distance scale. In fact, the relativistic cosmological fireball
model of M\'esz\'aros \& Rees (1993) predicts that the GRB spectra
have high-energy tails in general (M\'esz\'aros, Rees, \& Papathanassiou 1994).
Nevertheless, numerous array and Cerenkov telescope
burst searches at high energies have
only provided upper limits; {\it e.g.}  Kieda {\it et al.} (1996) (CASA-MIA),
Alexandreas {\it et al.} (1994) (CYGNUS), Aglietta {\it et al.} (1992) (EAS-TOP),
Krawczynski {\it et al.} (1996),
Funk {\it et al.} (1996) (HEGRA), Allen {\it et al.} (1995) (MILAGRO),  
Amenomori {\it et al.} (1995) (Tibet), 
and Connaughton {\it et al.} (1995) (Whipple).

The purpose of this work is to provide quantitative predictions of the 
expected burst rate  
as a function of photon energy accounting for cosmic absorption
and to compare the result with experimental limits.
We will start with the computation of the gamma-ray horizon
and show templates of absorbed spectra. In a second step,
we use the gamma-ray horizon to determine the number of 
bursts at a given detection threshold. Finally,
we determine the limiting sensitivity required 
for a detection at a given threshold gamma-ray energy and compare
this with the sensitivity of current experiments.
  
\section{Gamma-ray absorption from 10~GeV to 100~TeV}
Gamma-rays of energy $E$
propagating from a distant source at redshift $z_\circ$
toward a terrestrial observer
can be absorbed by inelastic interactions with low energy photons
of present-day energy $\epsilon$
from an isotropic diffuse background radiation field.  The dominant
process is pair creation $\gamma_E+\gamma_\epsilon
\rightarrow \rm e^++e^-$.  The threshold energy for this process
is given by
\begin{equation}
\label{eps}
\epsilon_{\rm th}={2(m_{\rm e}c^2)^2\over E(1-\mu)(1+z)^2}\, ,
\end{equation}
where $\mu=\cos\theta$ denotes the cosine of the scattering
angle.  Since the
soft photon density varies strongly with energy for 
typical radiation fields, the optical
depth $\tau_{\gamma\gamma}$ must also vary reflecting the
number density of target photons at the resonant energy $\propto E^{-1}$.
The pair creation cross section is given by
\begin{equation}
\label{sigma}
\sigma_{\gamma\gamma}={3\sigma_{\rm T}\over 16}
(1-\beta^2)\left[2\beta(\beta^2-2)+(3-\beta^4)\ln\left(1+\beta\over
1-\beta\right)\right]{\rm cm^2}\, ,
\end{equation}
where $\beta=\sqrt{1-1/\gamma^2}$ with $\gamma^2=\epsilon/\epsilon_{\rm th}$,
and where $\sigma_{\rm T}$ denotes the Thomson cross section.
For the computation of the optical depth we use the geodesic radial displacement
function 
$dl/dz={c\over H_\circ}[(1+z)E(z)]^{-1}$ where $E(z)$ is given by
equation (13.3) in Peebles (1993). For a cosmological model with $\Omega=1$
and $\Lambda=0$, the function $E(z)$ simplifies to $(1+z)^{3/2}$.
Hence, we obtain the optical depth
\begin{eqnarray}
\label{tau}
\tau_{\gamma\gamma}(E,z_\circ)=
\int_0^{z_\circ}dz{dl\over dz}\int_{-1}^{+1}d\mu{1-\mu\over 2}
\int_{\epsilon_{\rm th}}^\infty
d\epsilon n_{\rm b}(\epsilon)(1+z)^3\sigma_{\gamma\gamma}(E,\epsilon,\mu,z)
\nonumber\\
={c\over H_\circ}\int_0^{z_\circ}dz
(1+z)^{1/2}\int_{0}^{2}dx{x\over 2}\int_{\epsilon_{\rm th}}^\infty
d\epsilon n_{\rm b}(\epsilon)\sigma_{\gamma\gamma}(E,\epsilon,x-1,z)
\end{eqnarray}
for a nonevolving present-day background density $n_{\rm b}$,
i.e., 
$n_{\rm b}'(z,\epsilon')d\epsilon'
=(1+z)^3n_{\rm b}(\epsilon)
d\epsilon$, where the dash indicates comoving frame quantities. 
At some time in the past, the background photon density
was built up by bursts of massive star formation
during the era of galaxy formation.  As a result, the photon
density evolution should be more shallow than $\propto (1+z)^3$
at this epoch, and the corresponding $\gamma$-ray absorption
should be somewhat weaker.  We will discuss this effect only
qualitatively below.
The
shape of the present-day diffuse background density $n_{\rm b}(\epsilon)$
is obtained by averaging over various galaxy formation
models presented by MacMinn \& Primack (1996).
We multiplied this shape by a small factor to obtain
agreement with the
background density (including the contribution from the 2.7\ K microwave
background) estimated by Beichman \& Helou (1992)
in the far-infrared (FIR) (for a modest galaxy luminosity or density
evolution $\gamma=2$), and by Madau \& Phinney (1996) in the 
near-infrared (NIR) through UV range
(Fig.1).  The predicted background densities depend 
sensitively on which density or luminosity
evolution and maximum redshift are assumed and can vary by an order
of magnitude.   Hence, it is very important to tighten existing
constraints on the actual background density (Biller {\it et al.} 1995)
by obtaining
better galaxy luminosity functions from deep galaxy surveys
or by direct measurements of blazar gamma-ray (external!) turnover detections.

We integrate numerically the optical depth function
and solve for the gamma-ray horizon $\tau_{\gamma\gamma}(E_\circ,z_\circ)=1$.
Results 
are shown in Fig.2 for two values of the Hubble constant.
Templates of absorbed infinite power law spectra are shown in Fig.3
for various redshifts.

Note that the adopted diffuse radiation background does
not have a strong dust component which would be responsible
for TeV absorption of Mrk\,421, as suggested by de Jager, Stecker,
\& Salamon (1994).
Theoretical modeling of the broadband continuum spectrum of
Mrk\,421 over 18 orders of magnitude in frequency predicts
a cutoff due to {\it internal} gamma-ray absorption for Mrk421
(Mannheim {\it et al.} 1996) which could explain the lack of gamma-rays
above TeV observed by Whipple. This is in contrast to the
assumption by de Jager {\it et al.} (1994) 
of a straight power law for the intrinsic spectrum
extending well beyond TeV.
Further observations will show whether the cutoff is variable with flux as expected from
the intrinsic model or whether it is fixed as expected from absorption by
an anomalously high external dust component.

\section{Expected total number of bursts}
In our straw person's model, the bursts represent standard candles
out to a redshift of $z_{\rm max}=2.1$ (Cohen \& Piran 1995). 
Assuming no source evolution
and a galaxy density of $\sim 10^{-2}h^3$~Mpc$^{-3}$ 
($H_\circ=100h$~km~s$^{-1}$~Mpc$^{-1}$),
the observed BATSE brightness distribution can be matched if 
$\sim 10^{-6}$ bursts 
(each of energy $L_\circ = 7\times 10^{50}h^{-2}$\,ergs)
occur per year per galaxy. 
The number of bursts above a gamma-ray energy
$E_\circ$ is then given by the number of bursts 
in the cosmic volume enclosed by the gamma-ray horizon at that energy,
i.e. the hyperplane $\tau_{\gamma\gamma}(z_\circ, E_\circ)=1$.
The absorbed energy from bursts beyond the horizon is
re-emitted in other wave bands depending on the 
intergalactic magnetic
field (Aharonian, Coppi, \& V\"olk 1994), possibly with time delays
indicative of intergalactic magnetic fields (Plaga 1995).
For burst spectra as steep as those observed with EGRET
in the GeV range (Hurley {\it et al.} 1994), the ``pileup''
of absorbed energy below $E_\circ$ is negligible. 
The number of bursts per year at a given threshold energy $E_\circ$
and for a volume burst rate $n_\circ$ is then given by
\begin{equation}
\label{ne}
N(\ge E_\circ)=4\pi  n_\circ \left(c\over H_\circ\right)^3
\int_0^{z_\circ}dz {y(z)^2\over (1+z)E(z)}
 \end{equation} 
[see equation (13.61) of Peebles (1993), with an
extra factor (1+z) accounting for the redshift of the burst
rate]. 
The integral simplifies considerably for $\Omega=1$
and $\Lambda=0$. In this case, the angular size distance
$y(z)$ is given by
$y(z)=2[1-(1+z)^{-1/2}]$ and $E(z)=(1+z)^{3/2}$ yielding
\begin{equation}
N(\ge E_\circ)=16\pi  n_\circ \left(c\over H_\circ\right)^3
\left[{1\over 6}+{1\over 2(1+z_\circ)^{2}}-{2\over 3
(1+z_\circ)^{3/2}}\right].
\end{equation}
For the volume
burst rate $n_\circ= 6.9\times 10^{-9}h^{3}$\ Mpc$^{-3}$\ yr$^{-1}$
and
a BATSE trigger efficiency of $N_{\rm BATSE}/N=0.3$, we obtain
$N_{\rm BATSE}=300$\ yr$^{-1}$ integrating to $z_\circ=z_{\rm max}$.
The dependence on the Hubble constant enters
equation (\ref{ne}) only through the value of $z_\circ=z_\circ(E_\circ, H_\circ)$
and is therefore rather weak. Nevertheless, it is
an interesting possibility to determine the Hubble constant
by number counts of GRBs.
 
In Fig.4 we show
the number of bursts versus detection threshold demonstrating
the rapid shrinking of the observable universe with increasing
$E_\circ$ due to pair absorption. At energies of $\sim$10\ TeV,
the standard candle scenario predicts $2\sim 4$ ($h=0.5\sim 0.75$)
bursts per year.
 
\section{Limiting sensitivity}
The number of bursts which can be
{\it detected} above a given energy threshold depends also on the
particular instrumental sensitivity $F_\circ$, the spectral shape 
and the duration of the bursts.
Suppose we want to measure {\it all}
GRBs above an energy $E_\circ$ (within redshifts $z<z_\circ$), then   
what is the necessary flux sensitivity of our detector
to achieve this goal?
In the straw person's model, this problem is easy to solve.
For a given threshold $E_\circ$ the observed
flux obeys the relation
\begin{equation}
\label{fo}
F_\circ={L[\ge E_\circ(1+z_\circ)]\over 4\pi (c/H_\circ)^2
y(z_\circ)^2(1+z_\circ)^2}\, ,
\end{equation}
again referring to Peeble's (1993) notation. 
The standard candle luminosity normalized for a BATSE burst
duration $\delta t$ is given by 
\begin{equation}
L[\ge E_\circ(1+z_\circ)]=7\times 10^{50}h^{-2}\left(\delta t\over 1\ {\rm s}\right)^{-1}
\left(E_\circ\over E_{\rm b}\right)^{2-\alpha}{(1+z_\circ)^{2-\alpha}}\, ,
\end{equation}
adopting a high-energy power law
spectrum with photon index
$\alpha>2$ above an intrinsic break energy $E_{\rm b}\sim 1$~MeV.
With $\Omega=1$ and $\Lambda=0$, this corresponds to
\begin{equation}
F_\circ=1.7\times 10^{-7}{(1+z_{\circ})^{1-\alpha}
\over (\sqrt{1+z_\circ}-1)^2}\left(\delta t\over
1\ {\rm s}\right)^{-1}\left(E_\circ\over E_{\rm b}\right)^{2-\alpha}
\ {\rm ergs\ cm^{-2}\ s^{-1}}\, ,
\end{equation}
for the source flux at the distance of the gamma
ray horizon. All other sources will be within the horizon,
thus closer, and thus brighter. 
Although
the burst luminosity in the standard-candle picture is fixed
($\delta t\sim 1$~s), we leave the possibility open that the
high-energy tails persist longer than the typical BATSE bursts.
Hurley {\it et al.} (1994) indicate
spectral slopes $\alpha=2.2\sim3.7$ and extended
or delayed durations
of 30s $\sim$ 90 minutes for EGRET detected bursts.

In Fig.5 we plot the limiting sensitivity for $\alpha=2.2$ and
$\alpha=2.7$ and for high-energy tail durations extended relative
to the BATSE bursts by factors 1 and 100.
For comparison, we show current experimental limits
in the same plot demonstrating that they can in principle
detect all bursts. The apparently paradoxical increase of
the limiting flux with energy reflects the rapid shrinking
of the gamma-ray horizon: The higher the energy, the closer,
and thus the brighter, the bursts from the gamma-ray horizon are.
More pessimistically, 
if the intrinsic spectra were steeper than the EGRET
bursts, the sensitivity of current 
experiments is insufficient to detect all of them.
In view of the very small {\it total} burst
rates at high energies (see Fig.4), this would practically comply with 
zero expected event rates.

\section{Conclusions}
We estimated the expected gamma-ray burst rate as a function
of threshold energy.  
The straw person's scenario
considers standard candles at cosmological distances
emitting an intrinsically unabsorbed 
high-energy power law spectrum. With these simplifying assumptions,
we predict a maximum of $20\sim 40$ bursts per year in the TeV range,
roughly an order of magnitude less above 10\ TeV, and practically
none at $100$\ TeV.  
The perhaps more realistic case of a steep power law luminosity
distribution of the bursts does not significantly 
affect the maximum redshift
$z_{\rm max}$ of their cosmological distribution
(Cohen \& Piran 1995), and therefore it does not change the
expected burst rates.
An enhancement of the burst rate in the $10\sim 100$~GeV range by factors
of $\sim 2$ is possible if evolution of the background density
is important at redshifts $\sim 1$.
The sensitivities of current experiments
are sufficient to detect bursts if their spectra are not
steeper than $\alpha\sim 2.7$.
Taking into account sky exposure and
triggering efficiency (see Tab.\, 1), we rule out
the possibility of cosmological
burst detection by CYGNUS, EAS-TOP and CASA-MIA (and similar experiments
with thresholds at $\sim100$\ TeV).  
Due to its low threshold,
MILAGRO or future low-threshold 
air $\check{\rm C}$erenkov telescopes
could at most detect $\sim 10$ events per year.
The Whipple Observatory, the Tibet air shower array,
and HEGRA have very low expected detection rates 
($0.01\sim 1$ events per year).   
\begin{table}
\caption{Gamma-ray burst rates predicted for various experiments ($\Omega=1$, $\Lambda=0$,
$h=0.5$). Note that
a Hubble constant $h=0.75$
would increase all rates by a factor $\sim 2$. }
\begin{minipage}{\textwidth}
\renewcommand{\footnoterule}{\rule{0pt}{0pt}}
\renewcommand{\thefootnote}{\arabic{mpfootnote}}
\begin{tabular}{llll} \hline
Experiment & Trigger    & Energy    & GRB Rate \\
           & Efficiency & Threshold (TeV)& (year$^{-1}$) \\ \hline \hline
CASA-MIA   & 0.05       & 100       &  0.0003       \\
CYGNUS     & 0.05       & 100       &  0.0003        \\
EAS-TOP    & 0.05       & 100       &  0.0003       \\
HEGRA-Scintillators& 0.05      & 25        &  0.02        \\
CASA-DICE   & 0.0004    & 20        &  0.0002        \\
HEGRA-AIROBICC& 0.006   & 13        &  0.01        \\
MILAGRO    & 0.05       & 0.2       &  10         \\
Tibet      & 0.05       & 10        &  0.1        \\
Whipple\footnote{in order to start follow-up observations slewing the \v{C}erenkov
telescope can
  take up to 1h}    
	   & 0.006      & 0.3       &  1.0        \\ \hline
\end{tabular}
\end{minipage}
\end{table}
If the burst delays at high
energies are shorter than the telescope slewing time, the 
expected burst rate for air $\check{\rm C}$erenkov telescopes
would be down by $\sim 0.01$ owing to their small field of view.
The major uncertainties of the 
theoretical expectation values lie in the cosmic IR-to-UV photon
density ($\sim 10$ in the FIR), the Hubble constant (factor $< 2$)
and the effect of forward cascading in the intergalactic medium
which counteracts the pair absorption depending on the intergalactic
magnetic field strength
(Protheroe \& Stanev 1993; Aharonian, Coppi, \& V\"olk 1994; Plaga 1995).
A low background density, a large Hubble constant 
and weak intergalactic magnetic fields would act to
increase the
high-energy burst detection rate.  On the other hand, the high
infrared background inferred by de Jager, Stecker\, \& Salamon (1994)
and by Dwek \& Slavin (1994) assuming TeV absorption of Mrk\,421 would
further reduce the number of bursts above $\sim 1$~TeV by a factor
of $\sim 10$.
In any cosmological scenario, pair absorption by diffuse background
radiation leads to a rapidly decreasing number of bursts above
$\sim 10$~GeV. By contrast,
bursts originating in a galactic halo would not be affected by
absorption which occurs only on cosmic scales.
Halo bursts should therefore be much more numerous than cosmological
bursts at high energies, as has been pointed out
by Alexandreas {\it et al.} (1994) and Hurley (1996).  
In fact, if one clings to the assumption
of infinite power law tails,
the nondetection of GRBs by current experiments is in 
agreement with the straw person's scenario of a cosmological origin
of bursts, but it would require intrinsic spectral cutoffs below
$\sim 100$\ GeV in halo models.   
On the other hand, just a few burst detections above $\sim$50\ TeV
would render cosmological models highly unlikely.
If intrinsic cutoffs would be important, it is interesting to note
that the predictions for halo and cosmological burst models are  
again quite different.
In halo models, the brighter bursts should have cutoffs at lower
energies due to their higher compactness, whereas cosmological models
predict the opposite, since the bright sources would be near sources
with less external absorption. This crucial test could be performed
by future missions that target the favorable energy range $10\sim100$~GeV.

\section{Acknowledgements}
\noindent
We thank Matthew Baring, Trevor Weekes 
and our referee Ocker de Jager for their comments.
This work was supported in part by NASA under grant NAGW 5-1578.
D.H. and K.M. are grateful for hospitality and support 
(under NSF grant PHY94-07194) 
during their stay at the ITP, where this collaboration began.
B.F. is supported by the BMBF, Germany, under contract 05 2WT164.
K.M. acknowledges travel support by the Deutsche Forschungsgemeinschaft
under grant Ma~1545/5-1.

\newpage
\section{References}

\noindent
\hangindent=0.4cm
Aglietta, M., et al. 1992, Il Nuovo Cimento, 15C, 441 

\noindent
\hangindent=0.4cm
Aharonian, F.A., Coppi, P.S., \& V\"olk, H.J. 1994, ApJ, 423, L5

 \noindent
\hangindent=0.4cm
Alexandreas, D.E., et al.  1994, ApJ, 426, L1

 \noindent
\hangindent=0.4cm
Allen, G. E., et al. 1995, Proc. 24th Int. Cosmic Ray Conf. (Rome), 3, 516

 \noindent
\hangindent=0.4cm
Amenomori, M. et al. 1995, Proc. 24th  Int. Cosmic Ray Conf.  (Rome), 2,  112

 \noindent
\hangindent=0.4cm
Beichman, C.A., \& Helou, G.  1991, ApJ, 370, L1

 \noindent
\hangindent=0.4cm
Biller, S.D., et al.  1995, ApJ, 445, 227

 \noindent
\hangindent=0.4cm
Blandford, R.D., \& Levinson, A. 1995, ApJ, 441, 79

 \noindent
\hangindent=0.4cm
Briggs, M. S.,  et al. 1996, ApJ, 459, 40

 \noindent
\hangindent=0.4cm
Cohen, E., \& Piran, T. 1995, ApJ, 444, L25

 \noindent
\hangindent=0.4cm
Connaughton, V., et al. 1995, Proc. 24th Int. Cosmic Ray Conf. (Rome), 2, 96

 \noindent
\hangindent=0.4cm
de Jager, O.C., Stecker, F.W., \& Salamon, M.H. 1994, Nature, 369, 294

 \noindent
\hangindent=0.4cm
Dingus, B., et al. 1995, ApSS, 231, 187

 \noindent
\hangindent=0.4cm
Dwek, E., \& Slavin, J. 1994, ApJ, 436, 696

 \noindent
\hangindent=0.4cm
Funk, B., et al. 1996, Proc. 3rd Huntsville Symp. on
            GRBs, ed. C. Kouveliotou, M. Briggs, \& G.J. Fishman,
	    (New York: AIP), in press

 \noindent
\hangindent=0.4cm
Hartmann, D., Briggs, M.S., \& Mannheim, K. 1996 ApJ, in press

\noindent
\hangindent=0.4cm
Hurley, K.,  et al. 1994, Nature, 372, 652

 \noindent
\hangindent=0.4cm
Hurley, K., 1996,  Space Sci. Rev., 75, 43

 \noindent
\hangindent=0.4cm
Kieda, D., et al. 1996, Proc. 3rd Huntsville Symp. on
            GRBs, ed. C. Kouveliotou, M. Briggs,\& G.J. Fishman,
	    (New York: AIP), in press
            
 \noindent
\hangindent=0.4cm
Krawczynski, H., et al. 1996, Proc. 3rd Huntsville Symp. on
            GRBs, ed. C. Kouveliotou, M. Briggs, \& G.J. Fishman,
	    (New York: AIP), in press
            
 \noindent
\hangindent=0.4cm
Kolatt, T., Dekel, A., \& Lahav, O. 1995, MNRAS, 275, 797

 \noindent
\hangindent=0.4cm
MacMinn, D., \& Primack, J. 1996, Space Sci. Rev., 75, 413

 \noindent
\hangindent=0.4cm
Madau, P., \& Phinney, E.S. 1996, ApJ, 456, 124

 \noindent
\hangindent=0.4cm
Mannheim, K. 1993, A\&A, 263, 267

 \noindent
\hangindent=0.4cm
Mannheim, K., Westerhoff, S., Meyer, H., \& Fink, H.-H. 1996, A\&A, in press 

 \noindent
\hangindent=0.4cm
Meegan, C. A., et al. 1995, ApJ, 446, L15

 \noindent
\hangindent=0.4cm
M\'esz\'aros, P., \& Rees, M.J. 1993, ApJ, 405, 278

 \noindent
\hangindent=0.4cm
M\'esz\'aros, P., Rees, M.J., \& Papathanassiou, H. 1994, ApJ, 432, 181

 \noindent
\hangindent=0.4cm
Peebles, P.J.E. 1993,  Principles of Physical Cosmology, 
(Princeton: Univ. Press)

 \noindent
\hangindent=0.4cm
Plaga, R. 1995, Nature, 374, 430

 \noindent
\hangindent=0.4cm
Protheroe,  R. J.,  \& Stanev, T. 1993, MNRAS, 264, 191

 \noindent
\hangindent=0.4cm
Stecker, F.W., de Jager, O.C., \& Salamon, M.H. 1992, ApJ, 390, L49

 \noindent
\hangindent=0.4cm
Tegmark, M., Hartmann, D.H.,  Briggs, M.S., \& Meegan, C.A.
1996, ApJ, in press

 \noindent
\hangindent=0.4cm
Tully, R.B. 1992, ApJ, 388, 9
\newpage

\centerline{\bf Figure Captions}
\vskip.5cm

\noindent
{\bf Fig.1:}
Spectral energy distribution
of the diffuse background radiation.
Components are the 2.7~K cosmic microwave background (CMB),
dust and stellar light from galaxies.
{\it Solid line:} FIR-to-UV background adopted in the
present work (which closely resembles the spectrum obtained
by averaging over the
various cold + hot dark matter models of galaxy formation
of MacMinn \& Primack 1996).  {\it Straight line segment:}
FIR
background photon density inferred 
by Dwek \& Slavin (1994) assuming
TeV absorption for Mrk421. {\it Limits:}  Experimental upper limits
on the background IR density obtained by Biller {\it et al.} (1995).
\vskip0.15cm

\noindent
{\bf Fig.2}
Gamma-ray horizon $\tau_{\gamma\gamma}=1$ for the diffuse background 
radiation shown in Fig.1.   {\it Solid line: $h=0.5$. 
Dashed line:  $h=1.0$}.
\vskip0.15cm

\noindent
{\bf Fig.3}
Templates of absorbed $\alpha=2$ spectra at various redshifts
at which blazars have been detected
with EGRET ($h=0.5$).
\vskip0.15cm

\noindent
{\bf Fig.4}
GRB rate as a function of energy threshold.
{\it Solid line: $h=0.5$. Dashed line:  $h=1.0$}.
\vskip0.15cm

\noindent
{\bf Fig.5}
{\it Solid lines:}
Limiting sensitivity for detection of all
GRBs as a function of threshold energy  assuming
a high-energy tail spectral index $\alpha$ ($h=0.5$).
{\it Dashed lines:} Same for
high-energy tails with extended durations relative to the
BATSE bursts by a factor of 100.   {\it Symbols:}
current experimental flux limits for burst detection
(typically assuming $\delta t\sim 30$~s).

\newpage

\begin{figure}
\centerline{\psfig{figure=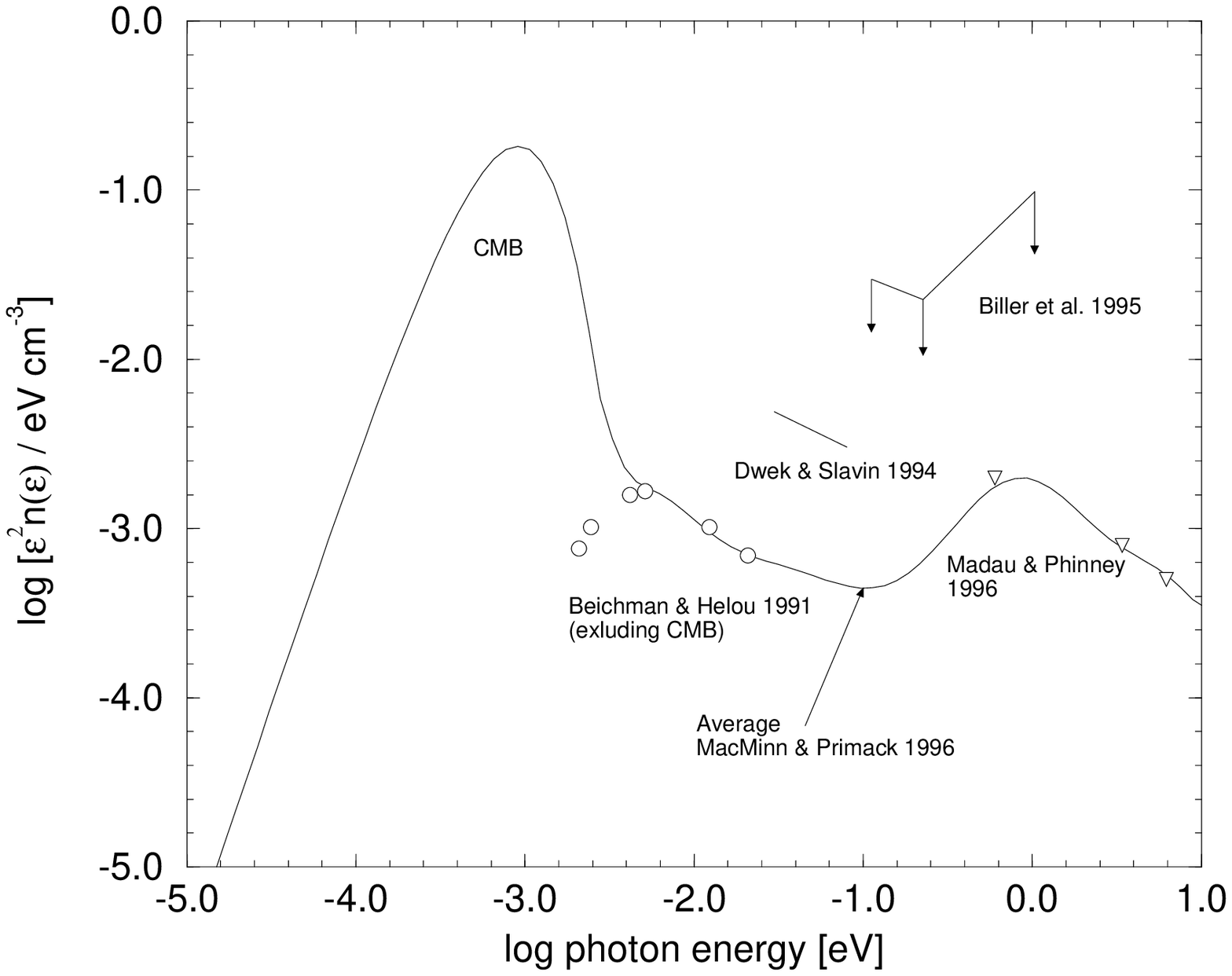,height=8cm,width=12cm}}
\caption[]{ 
}
\end{figure}

\begin{figure}
\centerline{\psfig{figure=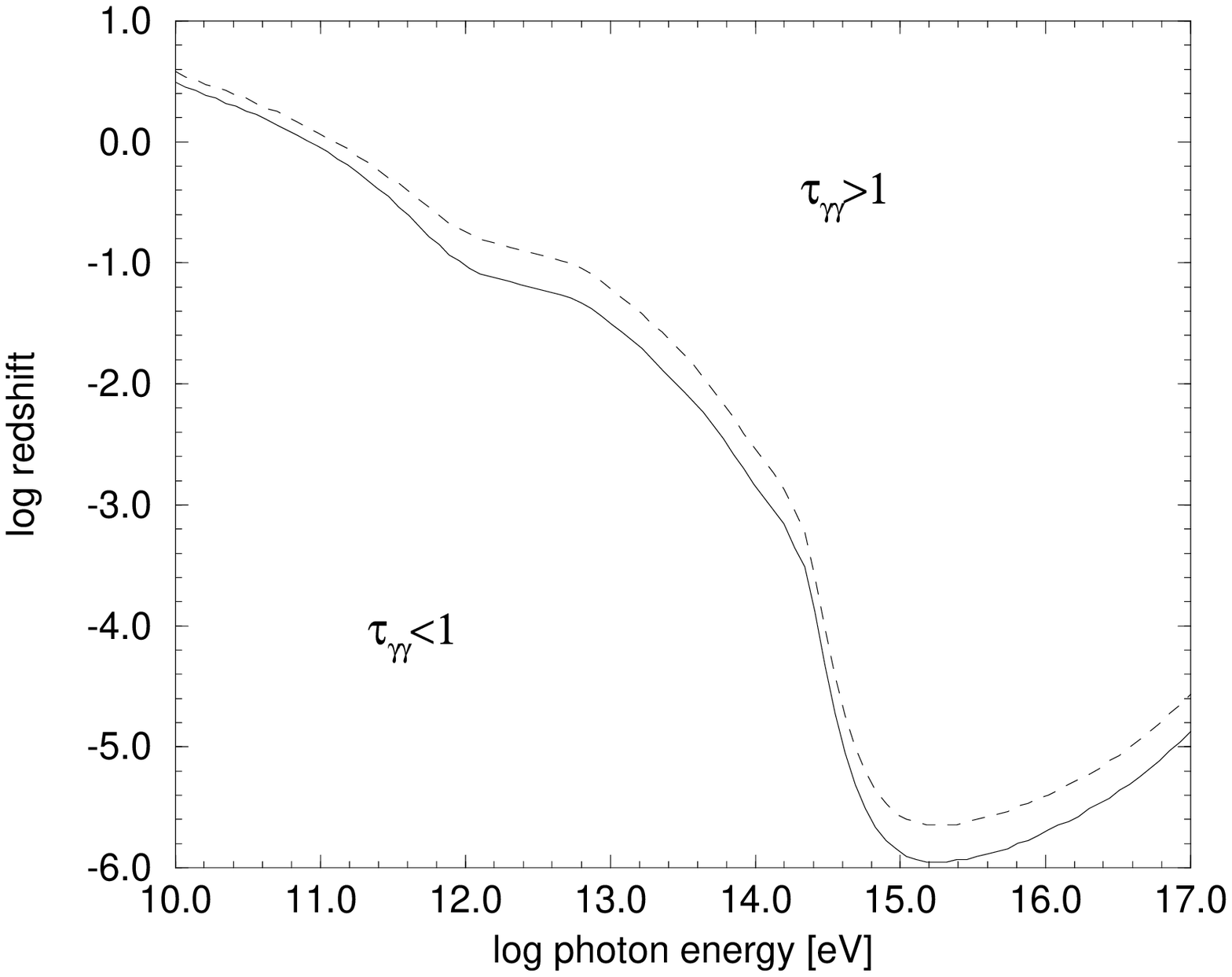,height=8cm,width=12cm}}
\caption[]{}
\end{figure}

\begin{figure}
\centerline{\psfig{figure=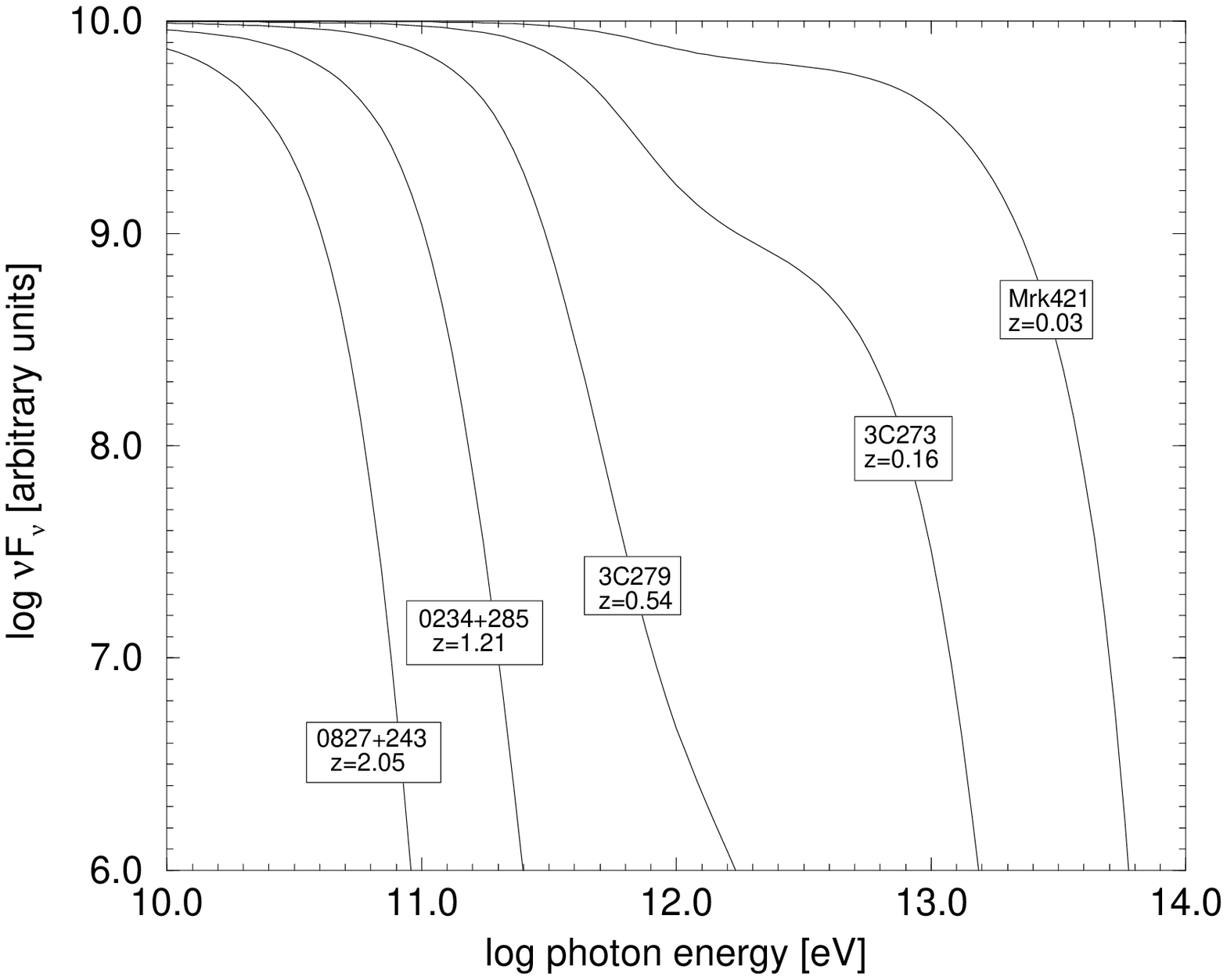,height=8cm,width=12cm}}
\caption[]{} 
\end{figure}

\begin{figure}
\centerline{\psfig{figure=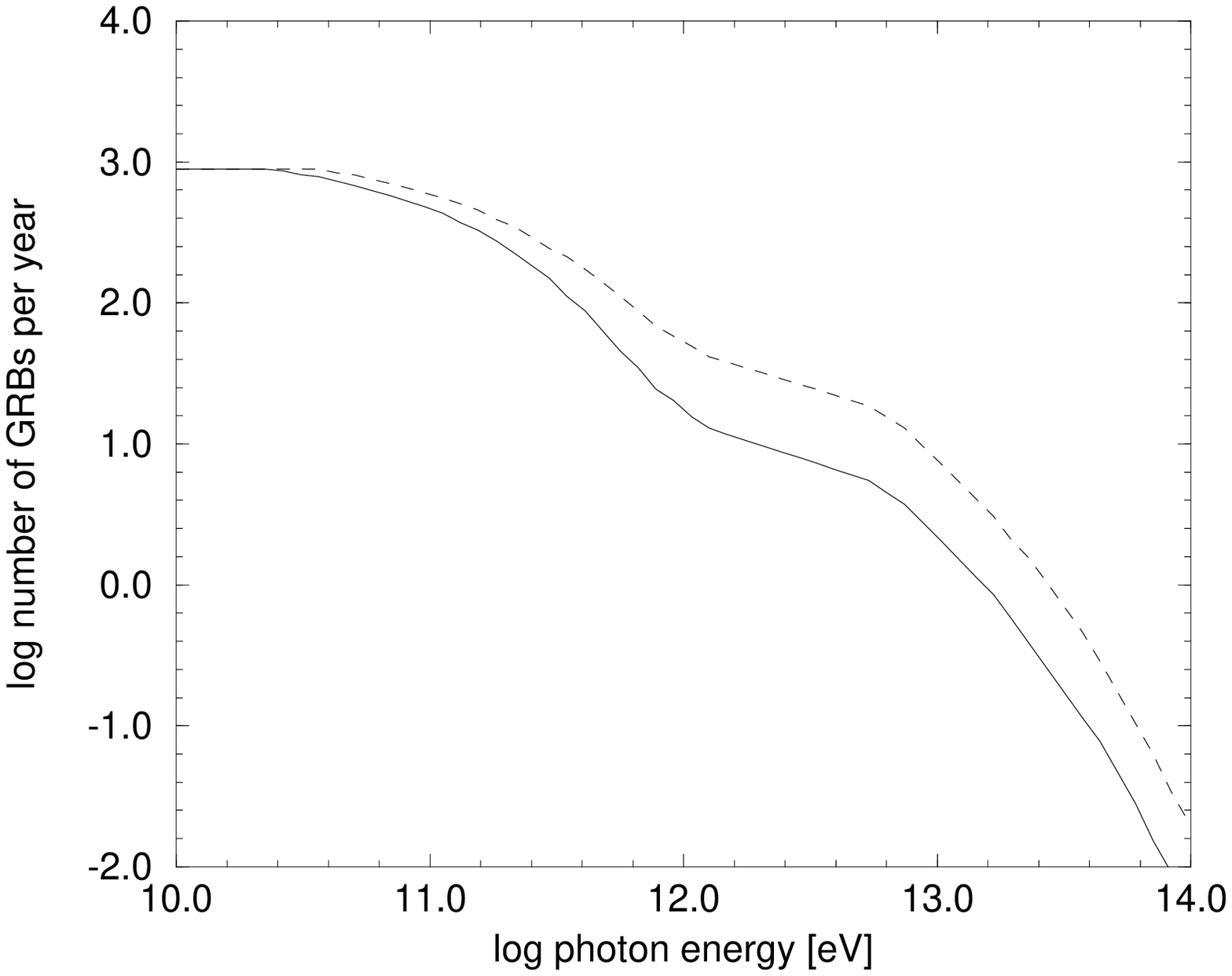,width=12cm,height=8cm}}
\caption[]{}
\end{figure}
 
\begin{figure}
\centerline{\psfig{figure=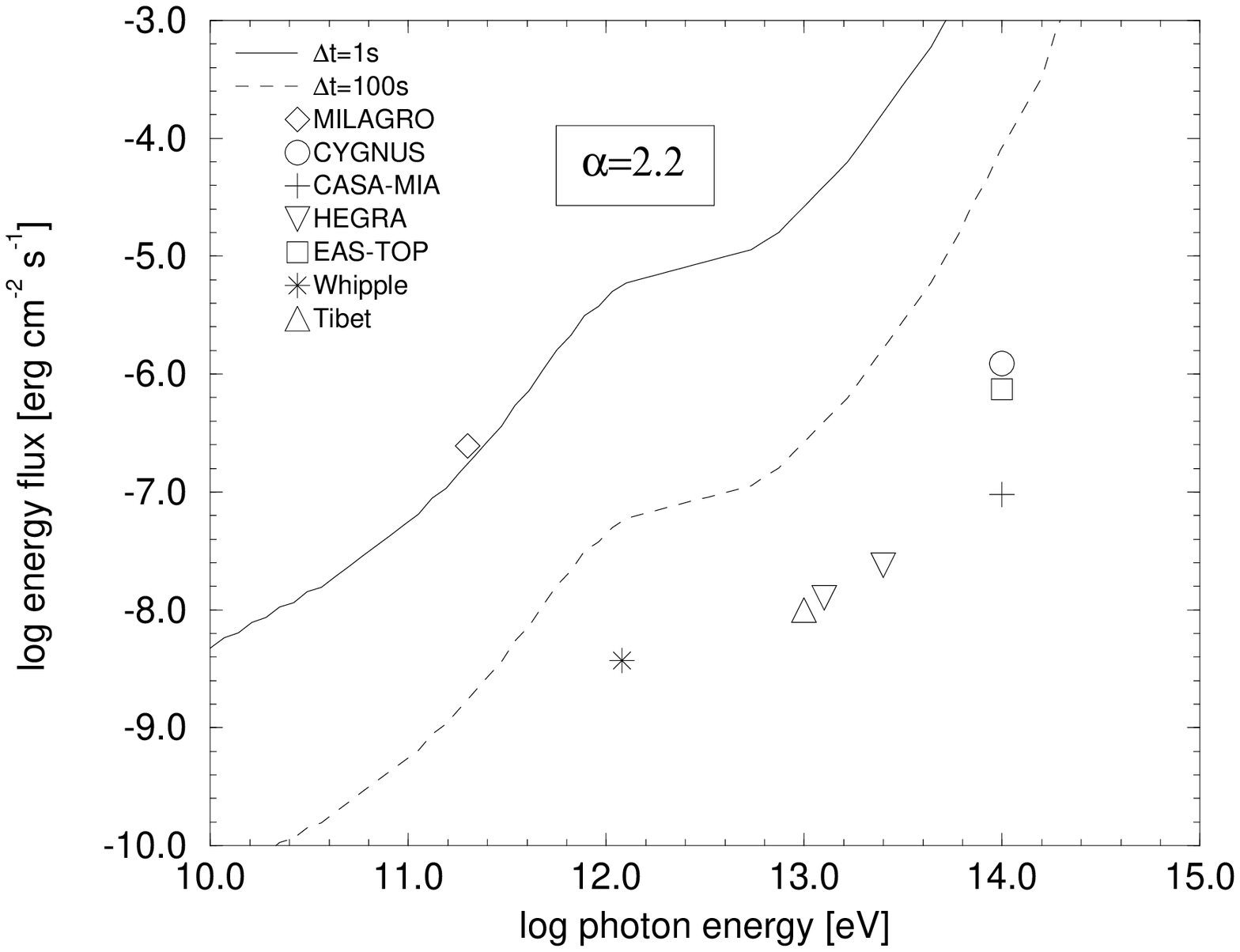,width=8cm,height=8cm}\psfig{figure=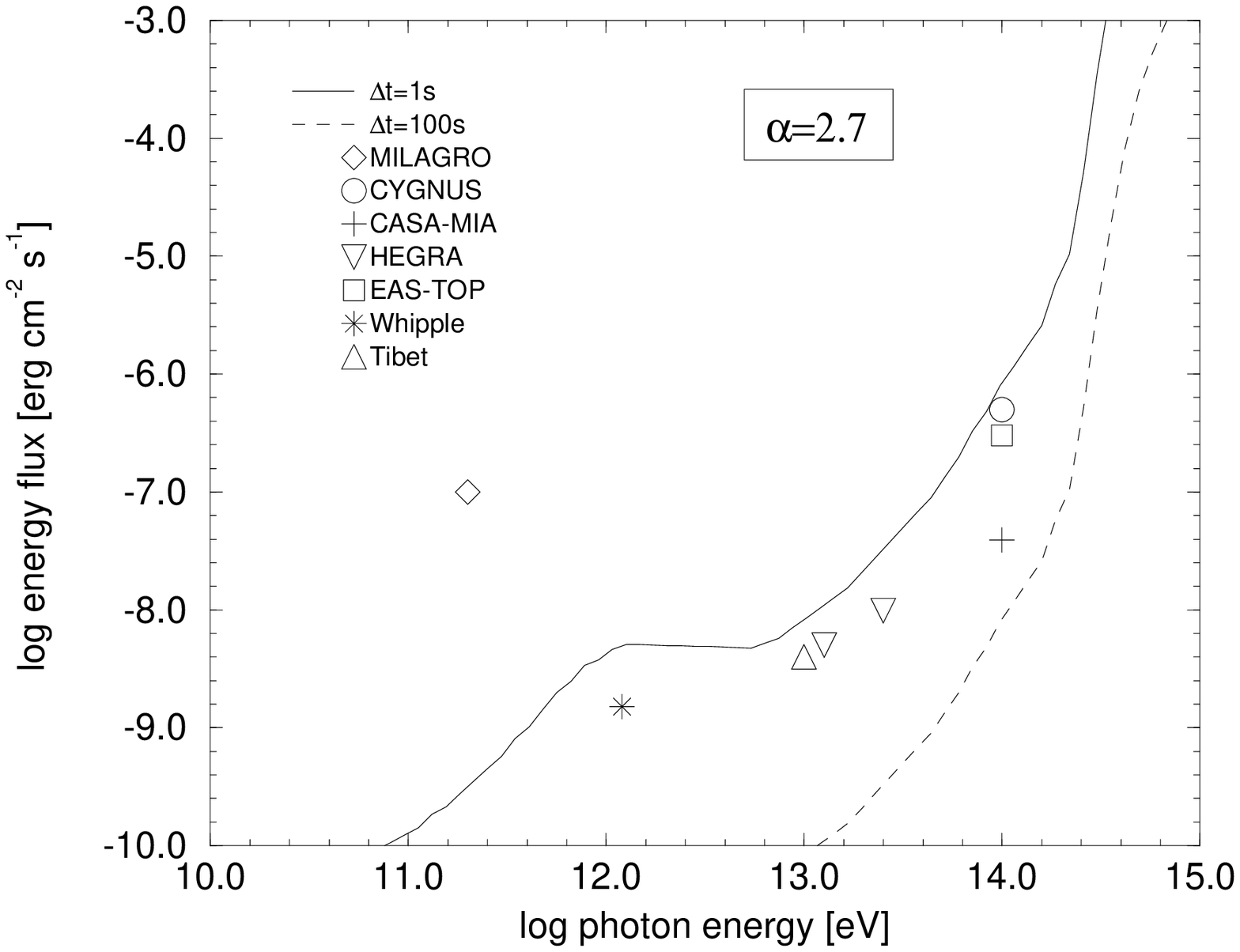,width=8cm,height=8cm}}
\caption[]{} 
\end{figure}

\end{document}